\documentclass[aps,prl,twocolumn,showpacs,superscriptaddress,groupedaddress]{revtex4} 
\usepackage{mathrsfs}
\usepackage{epsfig}
\usepackage{graphicx}
\usepackage{mathtools}
\usepackage{color}

\usepackage{url}
\usepackage{hyperref}

\begin{document}
\title{Planck Trispectrum Constraints on Primordial Non-Gaussianity at Cubic Order}

\author{Chang Feng\footnote{chang.feng@uci.edu}}
\affiliation{Department of Physics and Astronomy,
University of California, Irvine, CA 92697, USA }

\author{Asantha Cooray} 
\affiliation{Department of Physics and Astronomy,
University of California, Irvine, CA 92697, USA }

\author{Joseph Smidt}
\affiliation{XTD-IDA, Los Alamos National
Laboratory, Los Alamos, NM 87545}

\author{Jon O'Bryan}
\affiliation{Department of Physics and Astronomy,
University of California, Irvine, CA 92697, USA }

\author{Brian Keating}
\affiliation{Department of Physics and Astronomy,
University of California, San Diego, CA }

\author{Donough Regan}
\affiliation{Astronomy Centre, University of Sussex, Falmer, Brighton BN1 9QH, UK}

\begin{abstract}
Non-Gaussianity of the primordial density perturbations provides an important measure to constrain models of inflation.
At cubic order the non-Gaussianity is captured by two parameters $\tau_{\rm NL}$ and $g_{\rm NL}$ that determine the amplitude of the
density perturbation trispectrum.
Here we report measurements of the kurtosis power spectra of the cosmic microwave background (CMB) temperature as mapped by Planck
by making use of correlations between square temperature-square temperature and cubic temperature-temperature anisotropies.
In combination with noise simulations, we find the best joint estimates to be
 $\tau_{\rm{NL}}=0.3 \pm 0.9 \times 10^4$ and $g_{\rm{NL}}=-1.2 \pm 2.8 \times 10^5$.  If $\tau_{\rm NL}=0$, we find $g_{\rm NL}=
-1.3\pm 1.8 \times 10^5$.
\end{abstract}

\maketitle

{\it Introduction.}---Existing cosmological data from cosmic microwave background (CMB) and large-scale structure (LSS) are fully consistent with a simple cosmological
model involving six basic parameters describing the energy density components of the universe, age, and the amplitude and spectral index
of initial perturbations. The perturbations depart from a scale-free power spectrum and are Gaussian.
These facts support inflation as the leading paradigm related to the origin of density perturbations~\cite{Guth:1980zm,Linde:1981mu,Albrecht:1982wi}.
 Under inflation 
a nearly exponential expansion stretched space in the first moments of the early universe and promoted microscopic quantum fluctuations to perturbations on 
cosmological scales today~\cite{GuthPi,Bardeen}. 
Moving beyond simple inflationary models with a single scalar field, models of inflation now involve
multiple fields and exotic objects such as branes that have non-trivial interactions.  
Such inflationary models produce a departure from Gaussianity in a model-dependent manner~\cite{Byrnes:2010em,Engel:2008fu,Chen:2009bc,Boubekeur:2005fj}.  
The amplitude of non-Gaussianity therefore is an important cosmological parameter that can
distinguish between the plethora of inflationary models~\cite{Komatsu:2009kd}. 

The first order non-Gaussian parameter, $f_{\rm NL}$, has been measured with increasing success using the bispectrum - the Fourier analog of
the  three-point correlation function of the CMB temperature.
Such studies have found $f_{\rm NL }$ to be consistent with zero~\cite{Yadav:2007yy,Smith:2009jr,Komatsu:2010fb,Smidt:2009ir}, with the 
strongest constraint coming from Planck given by $f_{\rm NL}=2.7 \pm 5.8$~\cite{planckFNL}. The inflationary model expectation is that $f_{\rm NL} \lesssim 1$
and a constraint at such a low amplitude level may be feasible in the future with large scale structure data and with 21-cm intensity fluctuations.
Alternatively, with the trispectrum or four point correlation function of CMB anisotropies~\cite{Hu:2001fa}, we 
can measure the second and third order non-Gaussian parameters $\tau_{\rm NL}$ and $g_{\rm NL}$. While these higher order parameters
generally lead to a trispectrum that has a lower signal-to-noise ratio than the bispectrum, there may be models in which the situation is reversed with the trispectrum dominating over the bispectrum contribution. An example of such a model
is an inhomogeneous end to thermal inflation discussed in Ref.~\cite{Suyama13}.

A previous analysis using WMAP data out to $\ell < 600$ using the
kurtosis power spectra involving two-to-two and three-to-one temperature correlations~\cite{Smidt:2010ra,munshi},
found $-7.4 < g_{\rm NL}/10^5 < 8.2$ and   $-0.6 < \tau_{\rm NL}/10^4 < 3.3$  at the 95\% confidence level (C.L.). Other measures of the WMAP trispectrum have been presented in~\cite{wmap5ng,gnlONLY,fergusson10,regan13}.
While the Planck data have been used to constrain $\tau_{\rm NL} < 2800$ at the 95\% C.L. such a constraint ignored the signal associated with $g_{\rm NL}$ \cite{planckFNL}. Using all of the Planck data, the expectation is that $g_{\rm NL}$ can be constrained with a 68\% CL uncertainty of $6.7 \times 10^4$ \cite{gnlONLY} with $\tau_{\rm NL}=0$, while $\tau_{\rm NL}$ can be constrained down to 560 if $g_{\rm NL}=0$ \cite{kogo2006}. Here we present an analysis of the Planck temperature anisotropy maps by making use of kurtosis power spectra
to constrain $\tau_{\rm NL}$ and $g_{\rm NL}$ jointly.

{\it Theory.}---  We begin the discussion with the temperature trispectrum defined as~\cite{OH} 
\begin{eqnarray}
\langle a_{l_1m_1}a_{l_2m_2}a_{l_3m_3}a_{l_4m_4}\rangle&=&\sum_{LM}(-1)^M\begin{pmatrix}
  l_1 & l_2 & L \\
  m_1 & m_2 & -M 
 \end{pmatrix}\nonumber\\
&&\begin{pmatrix}
  l_3 & l_4 & L \\
  m_3 & m_4 & M 
 \end{pmatrix}T^{l_1l_2}_{l_3l_4}(L) \, ,
\end{eqnarray}
where we have introduced the Wigner 3-$j$ symbol. The angular trispectrum, $T^{l_1 l_2}_{l_3 l_4}(L)$, can be further expressed in terms of sums of the products of Wigner 3-$j$ or 6-$j$ symbols times the so-called {\it reduced} trispectrum, $\mathcal{T}^{l_1 l_2}_{l_3 l_4}(L)$~\cite{Hu:2001fa}.

To derive the angular trispectrum given by $T^{l_1l_2}_{l_3l_4}(L)$ we assume that the curvature perturbations $\zeta$ of the universe generated by inflation follow
as:
\begin{equation}
\Phi(\textbf{x})=\Phi_G(\textbf{x})+f_{\rm{NL}}(\Phi_G^2(\textbf{x})-\langle\Phi^2_G(\textbf{x})\rangle)+g_{\rm{NL}}\Phi^3_G(\textbf{x}) \ .
\end{equation}
where the curvature perturbation $\zeta$ and the initial gravitational potential are related by $\Phi=(3/5)\zeta$ and $\tau_{\rm{NL}}=(6f_{\rm{NL}}/5)^2$.

We refer the reader to Ref.~\cite{kogo2006} for intermediate steps in our derivation. Using the above form the full trispectrum can be
reduced to two forms involving the two amplitudes $\tau_{\rm NL}$ (associated with $\Phi_G^2(\textbf{x})-\langle\Phi^2_G(\textbf{x})\rangle$ term
in above) and $g_{\rm{NL}}$ coming from $\Phi^3_G(\textbf{x})$. 

Defining $\mathcal{T}^{l_1l_2, (i)}_{l_3l_4}(L)=h_{l_1l_2L}h_{l_3l_4L}t^{l_1l_2,(i)}_{l_3l_4}(L)$, $i=1,2$~\cite{regan10}, where
\begin{equation}
h_{l_1l_2l_3}=\sqrt{\frac{(2l_1+1)(2l_2+1)(2l_3+1)}{4\pi}}\begin{pmatrix}
  l_1 & l_2 & l_3 \\
  0 & 0 & 0 
 \end{pmatrix},
\end{equation}
we find that the reduced trispectrum is
\begin{equation}
\mathcal{T}^{l_1l_2}_{l_3l_4}(L)=[\tau_{\rm{NL}}\mathcal{T}^{l_1l_2,(1)}_{l_3l_4}(L)+g_{\rm{NL}}\mathcal{T}^{l_1l_2,(2)}_{l_3l_4}(L)].
\label{signal}
\end{equation}
The two terms are
\begin{eqnarray}
t^{l_1l_2,(1)}_{l_3l_4}(L)&=&\tau_{\rm NL} \Big(\frac{5}{3}\Big)^2\int r_1^2dr_1r^2_2dr_2F_L(r_1,r_2)\nonumber\\
&&\alpha_{l_1}(r_1)\beta_{l_2}(r_1)\alpha_{l_3}(r_2)\beta_{l_4}(r_2) \, ,\label{tNLpart}
\end{eqnarray}
and
\begin{eqnarray}
t^{l_1l_2,(2)}_{l_3l_4}(L)&=&g_{\rm NL}\int r^2dr\beta_{l_2}(r)\beta_{l_4}(r)[\alpha_{l_1}(r)\beta_{l_3}(r)\nonumber\\
&&+\alpha_{l_3}(r)\beta_{l_1}(r)] \, .\label{gNLpart}
\end{eqnarray}
Here $
\alpha_l(r)=(2/\pi)\int k^2dk\Delta^{\rm{TT}}_l(k)j_l(kr)$ and $\beta_l(r)=(2/\pi)\int k^2dkP(k)\Delta^{\rm{TT}}_l(k)j_l(kr)$. 
The primordial curvature power spectrum is $k^3P(k)/(2\pi^2)=(3/5)^2A_s(k/k_0)^{n_s-1}$ with no ``running"\cite{planckparas}. 
Here $k_0$ is the pivot scale set at $0.05\rm{Mpc}^{-1}$. 
We use the public code~\footnote{\url{http://www.mpa-garching.mpg.de/~komatsu/CRL/nongaussianity/localform/}} to compute $\alpha_l(r)$, $\beta_l(r)$ 
and the temperature transfer function $\Delta_l^{\rm{TT}}(k)$. 

In the $\tau_{\rm{NL}}$ part, we define the function $F_L$ as 
\begin{equation}
F_L(r_1,r_2)=\frac{2}{\pi}\int k^2dkP(k)j_L(kr_1)j_L(kr_2).\label{Fexact}
\end{equation}
Following the efficient algorithm in~\cite{r-inte-4}, we define $\xi=r_2/r_1$, $x=kr_1$ and compress $r_1$ and $r_2$ into one dimension such that
\begin{equation}
F_L(\xi)=\frac{2}{\pi}r_1^{1-n_s}\lambda\int dx x^{n_s-2}j_L(x)j_L(tx),
\end{equation}
Here $\lambda=(3/5)^2(2\pi^2/k_0^3)A_sk_0^{4-n_s}$. We validate that this fast algorithm gives the same results as Eq.~\ref{Fexact}.

The first part of the trispectrum associated with $\tau_{\rm{NL}}$ approximates to $(5/3)^2C_L^{r_{\ast}}\sqrt{C_{l_1}C_{l_2}C_{l_3}C_{l_4}}$ at $L<100$. This is due to the fact that
the integrand peaks at $r=r_{\ast}$ 
and $C_l=\int r^2dr\alpha_l(r)\beta_l(r)$~\cite{ruth}. 
Here $r_{\ast}$ is the comoving distance at last scattering surface and $C_L^{r_{\ast}}=F_L(r_{\ast},r_{\ast})$. For the comparison with the data, however, we perform an exact calculation defined in Eqs.~\ref{tNLpart},~\ref{gNLpart}. 
The adaptive $r$-grid is used for the integration. 

The estimators of the connected trispectrum are constructed in Refs.~\cite{munshi,munshi1} and they are given by
\begin{equation}
K_L^{(2,2)}(\tau_{\rm{NL}},g_{\rm{NL}})=\frac{1}{2L+1}\sum_{l_1l_2l_3l_4}\frac{1}{2L+1}\frac{\mathcal{T}^{l_1l_2}_{l_3l_4}(L)\hat T^{l_1l_2}_{l_3l_4}(L)}{C_{l_1}C_{l_2}C_{l_3}C_{l_4}},
\label{k22}
\end{equation}
and
\begin{equation}
K_{l_4}^{(3,1)}(\tau_{\rm{NL}},g_{\rm{NL}})=\frac{1}{2l_4+1}\sum_{l_1l_2l_3L}\frac{1}{2L+1}\frac{\mathcal{T}^{l_1l_2}_{l_3l_4}(L)\hat T^{l_1l_2}_{l_3l_4}(L)}{C_{l_1}C_{l_2}C_{l_3}C_{l_4}}.
\label{k31}
\end{equation}
In Eqs.~\ref{k22},~\ref{k31}, the reduced trispectrum $\mathcal{T}^{l_1l_2}_{l_3l_4}(L)$ is evaluated at $\tau_{\rm{NL}}=1$ and $g_{\rm{NL}}=1$. The estimators $K_L^{(2,2)}$ and $K_L^{(3,1)}$ are parametrized by these two parameters. The $\hat T^{l_1l_2}_{l_3l_4}(L)$ denotes the full trispectrum from data or simulation.

\begin{figure}
\rotatebox{0}{\includegraphics[width=9.6cm, height=6cm]{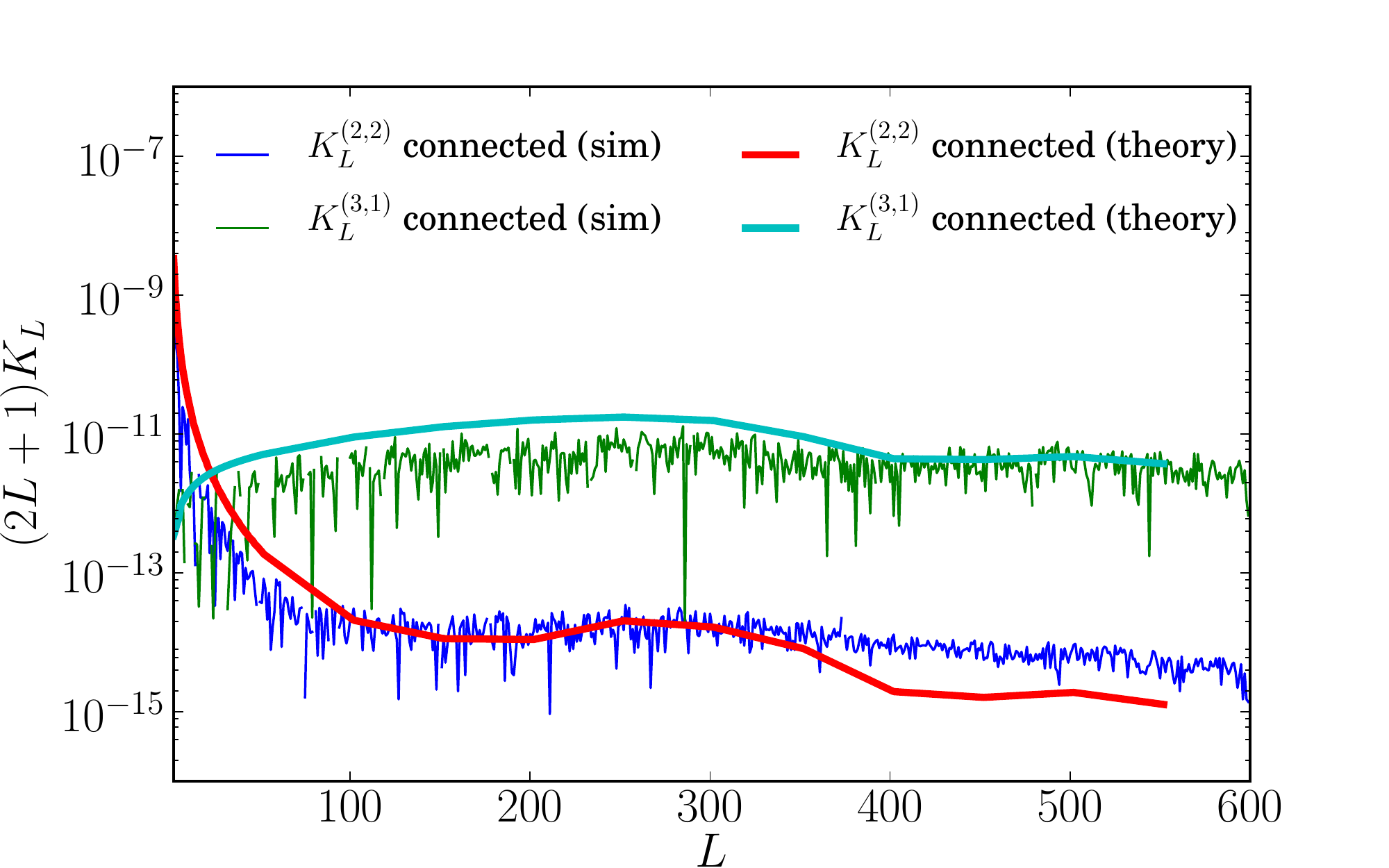}}
\caption{The estimator validation using WMAP simulations with $\tau_{\rm{NL}}=3600$.}
\label{Validation}
\end{figure}

In our analysis, $l_{\rm{min}}\le l_1,l_2,l_3,L\le l_{\rm{max}}$, $l_{\rm{min}}=2$ and $l_{\rm{max}}=1000$. 
The trispectrum 
computing time is proportional to $\mathcal{O}(l^4_{\rm{max}})$ at a single $L$. In order to make these calculations more efficient, we use Monte Carlo integration for $K_L^{(2,2)}$, i.e., replacing $\sum_{l_1=l_{\rm{min}}}^{l_{\rm{max}}}\sum_{l_2=l_{\rm{min}}}^{l_{\rm{max}}}\sum_{l_3=l_{\rm{min}}}^{l_{\rm{max}}}\sum_{l_4=l_{\rm{min}}}^{l_{\rm{max}}}$ by $V/N_{\rm{samples}}\sum_{\textbf{l}}$. The vector $\textbf{l}$(=$l_1,l_2,l_3,l_4$) is uniformly sampled from $[l_{\rm{min}},l_{\rm{max}}]^4$ and $V=(l_{\rm{max}}-l_{\rm{min}})^4$. For $K_L^{(3,1)}$, we restrict the diagonal elements within $2\le L\le 20$ and validate that a bigger upper bound negligibly modifies the trispectrum. The Wigner 3-$j$ symbols' intrinsic selection rule also helps reduce the computation time. With all these efficient algorithm, we can achieve a hour-level computation time, which is about three orders of magnitude faster than the brute-force calculation. We show the theoretical predictions of these estimators for the case in Fig.~\ref{Validation}
for a fixed set of $\tau_{\rm NL}$ and $g_{\rm NL}$ values for which non-Gaussian simulated maps are available.

From simulated and real data, spherical harmonic coefficients $a^{(\rm sim)}_{lm}$ and $a^{(\rm data)}_{lm}$ are computed by inverse spherical harmonic transformation (SHT). Then the two weighted maps are generated from definitions $A(r,\textbf{n})=\sum_{lm}\alpha_l(r)\tilde a_{lm}Y_{lm}(\textbf{n})$ , $B(r,\textbf{n})=\sum_{lm}\beta_l(r)\tilde a_{lm}Y_{lm}(\textbf{n})$ and $\tilde a_{lm}=a_{lm}/C_l$ where the angular power spectrum $C_l$ is inclusive of noise. $a_{lm}^{\rm{(data)}}$ is calculated by \textit{anafast} of Healpix which removes monopole and dipole. To correct the masking effect, we scale the masked modes $a^{(\rm sim)}_{lm}$ and $a^{(\rm data)}_{lm}$ by $1/\sqrt{f_{\rm{sky}}}$ to match the underlying temperature power spectrum. These masked modes are also beam- and pixel window-deconvolved. In the following text, we neglect ``$\textbf{n}$" for brevity. 

From $A$ and $B$ maps, we construct $C(r_1,r_2)=A(r_1)B(r_2)$. Then we make $C'_{lm}=F_L(r_1,r_2)C_{lm}(r_1,r_2)$ and $D(r_1,r_2)=C'(r_1,r_2)A(r_2)$. We can calculate four types of power spectra:
\begin{equation}
J_l^{\rm{ABA,B}}(r_1,r_2)=\frac{1}{2l+1}\sum_m D_{lm}(r_1,r_2)B^{\ast}_{lm}(r_2),
\end{equation}

\begin{eqnarray}
J_l^{\rm{AB,AB}}(r_1,r_2)&=&\frac{1}{2l+1}\sum_mF_l(r_1,r_2)\nonumber\\
&&[AB]_{lm}(r_1)[AB]^{\ast}_{lm}(r_2);
\end{eqnarray}
\begin{equation}
L_l^{\rm{ABB,B}}(r)=\frac{1}{2l+1}\sum_m[ABB]_{lm}(r)B^{\ast}_{lm}(r);
\end{equation}
and
\begin{equation}
L_l^{\rm{AB,BB}}(r)=\frac{1}{2l+1}\sum_m[AB]_{lm}(r)[BB]^{\ast}_{lm}(r).
\end{equation}

When all the power spectra are integrated along the line of sight, they become:
\begin{equation}
J_l^{\rm{ABA,B}}=\int r_1^2dr_1r_2^2dr_2J_l^{\rm{ABA,B}}(r_1,r_2);
\end{equation}
\begin{equation}
L_l^{\rm{ABB,B}}=\int r^2drL_l^{\rm{ABB,B}}(r);
\end{equation}
\begin{equation}
J_l^{\rm{AB,AB}}=\int r_1^2dr_1r_2^2dr_2J_l^{\rm{AB,AB}}(r_1,r_2);
\end{equation}
and
\begin{equation}
L_l^{\rm{AB,BB}}=\int r^2drL_l^{\rm{AB,BB}}(r).
\end{equation}

The trispectrum estimators 
\begin{equation}
K_L^{(2,2)}=\Big (\frac{5}{3}\Big )^2J_L^{\rm{AB,AB}}+2L_L^{\rm{AB,BB}},
\end{equation}
and
\begin{equation}
K_L^{(3,1)}=\Big (\frac{5}{3}\Big )^2J_L^{\rm{ABA,B}}+2L_L^{\rm{ABB,B}}
\end{equation}
are then constructed from the correlations associated with $A$ and $B$ maps that are either from data or simulations.

These estimators are applied to 143 GHz and 217 GHz temperature datasets, as well as the cross-correlation $143\times217$ GHz. For the cross correlation, the estimators are
\begin{eqnarray}
K_L^{(2,2)}(143\times 217)&=&\Big (\frac{5}{3}\Big )^2J_L^{\rm{A(143)B(217),A(143)B(217)}}\nonumber\\&+&2L_L^{\rm{A(143)B(217),B(143)B(217)}},
\end{eqnarray}
and
\begin{eqnarray}
K_L^{(3,1)}(143\times 217)&=&\Big (\frac{5}{3}\Big )^2J_L^{\rm{A(143)B(217)A(143),B(217)}}\nonumber\\&+&2L_L^{\rm{A(143)B(217)B(143),B(217)}}.
\end{eqnarray}

{\it Simulation Validation:}
To validate our estimates of the connected trispectra, we make non-Gaussian CMB signal simulations. The non-Gaussian maps for WMAP are publicly available~\footnote{\url{http://planck.mpa-garching.mpg.de/cmb/fnl-simulations/}} so we simulate maps with $n_{\rm{side}}=512$ and $l_{\rm{max}}=600$, and all 
the WMAP experimental settings, consistent with 5-year observations, are adopted. For the signal part, $a_{lm}=a^{\rm{G}}_{lm}+f_{\rm{NL}}a^{\rm{NG}}_{lm}$ and we choose $f_{\rm{NL}}=50$, i.e., $\tau_{\rm{NL}}=3600$
given the expected relation between $f_{\rm NL}$ and $\tau_{\rm NL}$, independent of the exact value of $g_{\rm NL}$. Note that the non-Gaussian
simulations we use assume $g_{\rm NL}=0$ and in a joint model fit to data we test this expectation.
 The WMAP 5-yr noises are then added in the signal simulations. The WMAP simulation is $T(\textbf{n})=\sum_{lm}b_lp_la_{lm}Y_{lm}(\textbf{n})+\sigma_0/\sqrt{N(\bf{n})}n_{\rm{white}}(\textbf{n})$. Here $\sigma_0$ and $N(\textbf{n})$ are provided by WMAP. The estimator of the connected trispectrum is $\hat K_L=1/4!(K_L-K^{\rm{Gaussian}}_L)$. In Fig.~\ref{Validation} 
we show that the average connected parts from 100 full-sky realizations are consistent with the theoretical calculations.

{\it Data Analysis and Results:} 
We use Planck 143 GHz and 217 GHz temperature maps for the present analysis. We use the foreground mask to remove the point sources and galactic emissions for both frequencies. The 217 GHz map cleaned after the $70\%$ foreground mask still contains visible emission around the galactic plane, so we use an extended mask to further cut the 217 GHz data around it. The resulting sky fractions for both maps become $73\%$ and $58\%$.
At 143 GHz, the map is convolved with a $7'$ Gaussian beam and has $45\mu {\rm K}{\ \rm arcmin}$ noise. At 217 GHz, it is $5'$ and $60\mu {\rm K}{\ \rm arcmin}$.
Following Ref.~\cite{plancklensing},
point sources (PS) and cosmic infrared background (CIB) are also included in simulated data. The power spectra for these two sources are $C_l^{\rm{PS}}=2\pi/3000^2$ and $C_l^{\rm{CIB}}=2\pi/(l(l+1))(l/3000)^{0.8}$, respectively. The foreground power at these frequencies are $C_l^{A\times B}=A^{\rm{PS}}_{A\times B}C_l^{\rm{PS}}+A^{\rm{CIB}}_{A\times B}C_l^{\rm{CIB}}$ with the parameters $A^{\rm{PS}}_{143\times 143}=64 \mu {\rm K}^2, A^{\rm{PS}}_{143\times 217}=43 \mu {\rm K}^2, A^{\rm{PS}}_{217\times 217}=57 \mu {\rm K}^2, A^{\rm{CIB}}_{143\times 143}=4 \mu {\rm K}^2, A^{\rm{CIB}}_{143\times 217}=14 \mu {\rm K}^2, A^{\rm{CIB}}_{217\times 217}=54 \mu {\rm K}^2$.
In addition, a $10\mu {\rm K}{\ \rm arcmin}$ white noise is added into the simulations. The data structure is expressed as $T(\textbf{n})=\sum_{lm}a_{lm}b_lp_lY_{lm}(\textbf{n})+n(\textbf{n})$ where $\textbf{n}$ is a direction on the sky, $b_l$ is the beam transfer function, $p_l$ is the pixel transfer function at $n_{\rm{side}}=2048$, and $n(\textbf{n})$ is the noise simulation. We use 100 signal and noise realizations from the FFP6 simulation set of the 
Planck collaboration~\cite{planckFFP6}. We use the best-fit cosmological parameters from ``Planck+WP+highL" \cite{planckparas}. Specifically, $\Omega_bh^2=0.022069$, $\Omega_ch^2=0.12025$, $\tau=0.0927$, $n_s=0.9582$, $A_s=2.21071\times10^{-9}$ at pivot scale $k_0=0.05\rm{Mpc}^{-1}$, and $H_0=67.15{\rm{km{\ } s^{-1}Mpc^{-1}}}$~\cite{planckparas}.

\begin{figure}
\rotatebox{0}{\includegraphics[width=9.6cm, height=6cm]{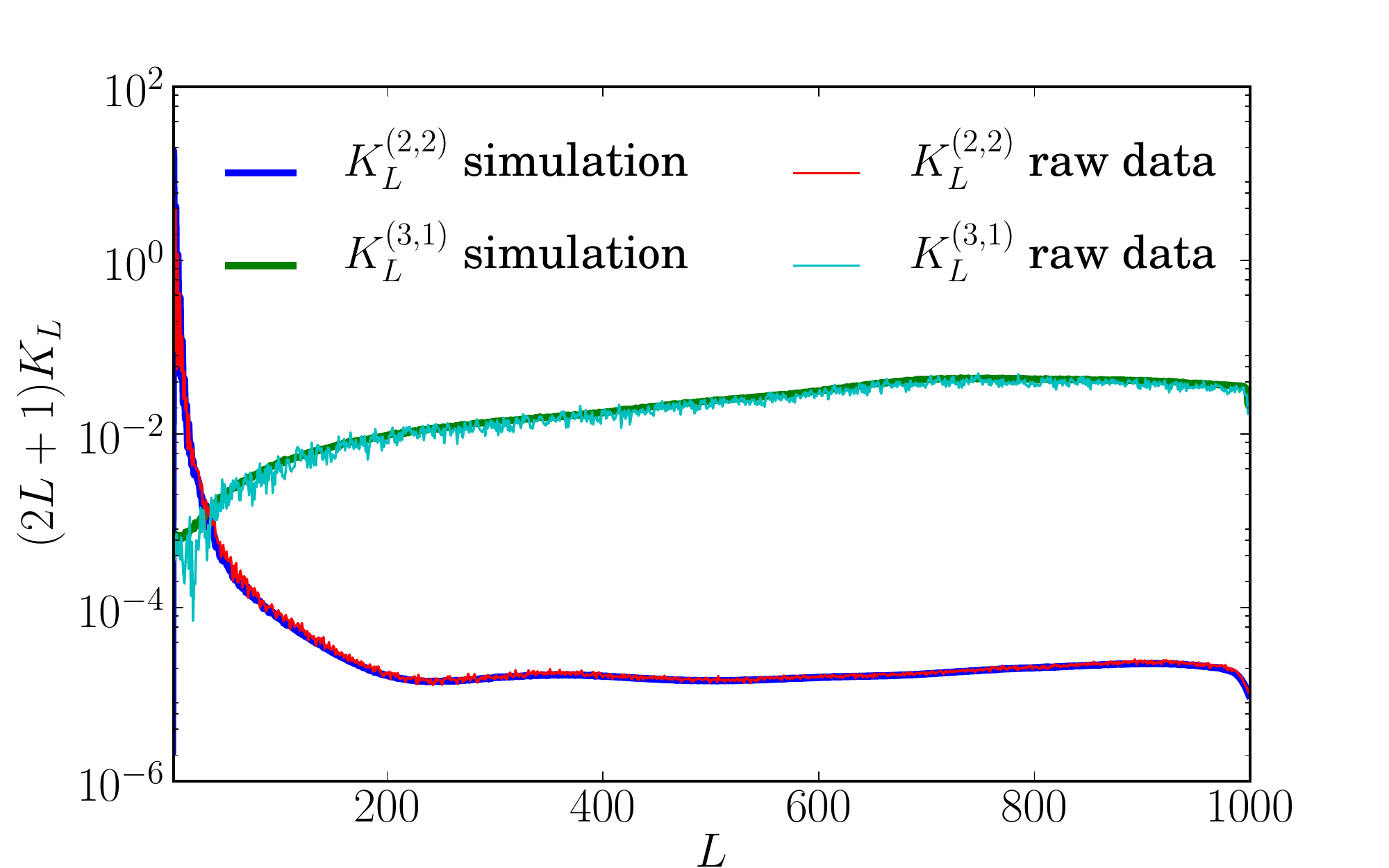}}
\rotatebox{0}{\includegraphics[width=9.6cm, height=6cm]{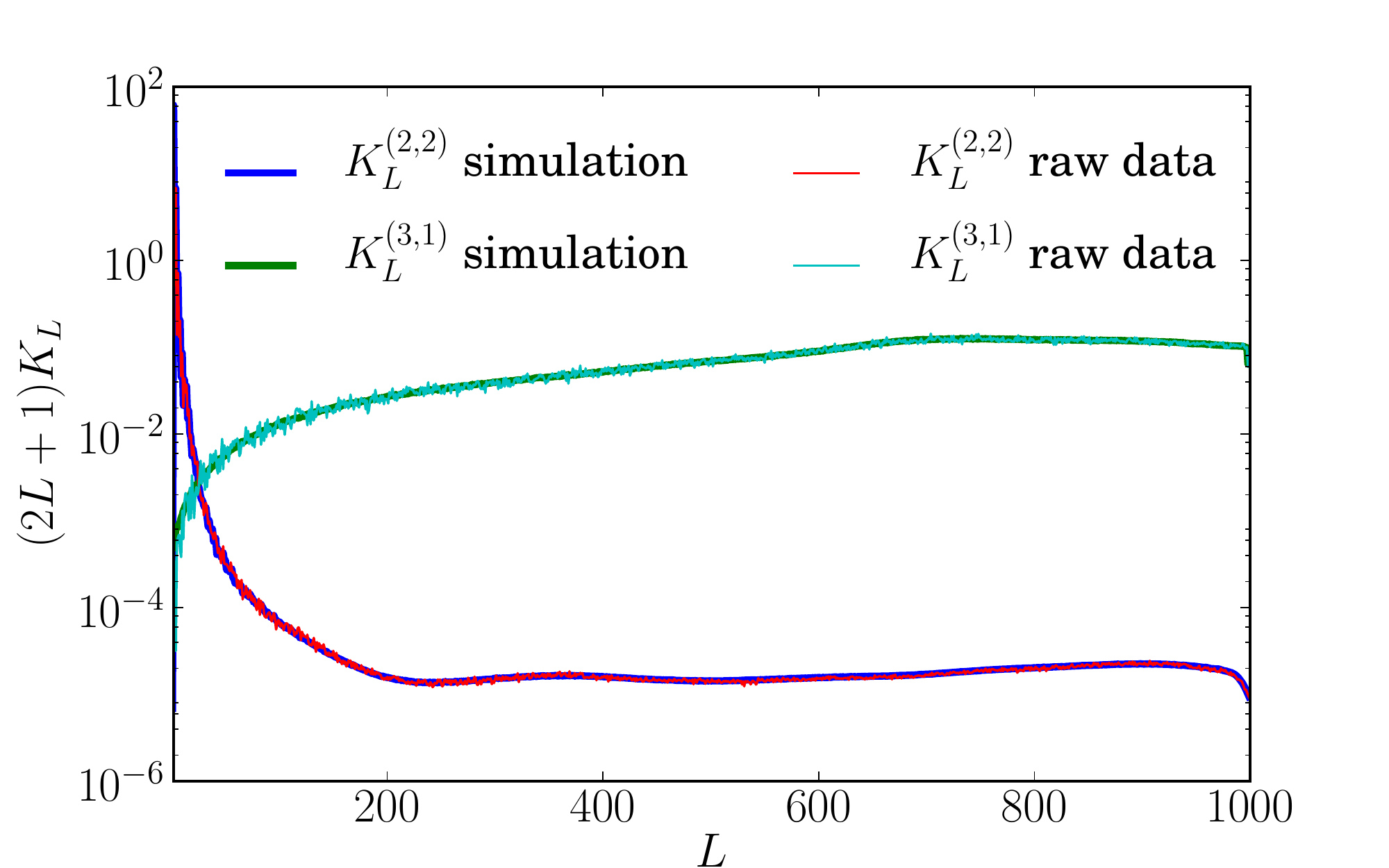}}
\caption{The raw trispectra calculated from Planck data and simulations for $143\times 143$ GHz (top) and $143\times 217$ GHz (bottom). 
In both plots Gaussian bias dominates the raw signal.}
\label{RawTrispectrum}
\end{figure}

\begin{figure}
\rotatebox{0}{\includegraphics[width=9.1cm, height=8cm]{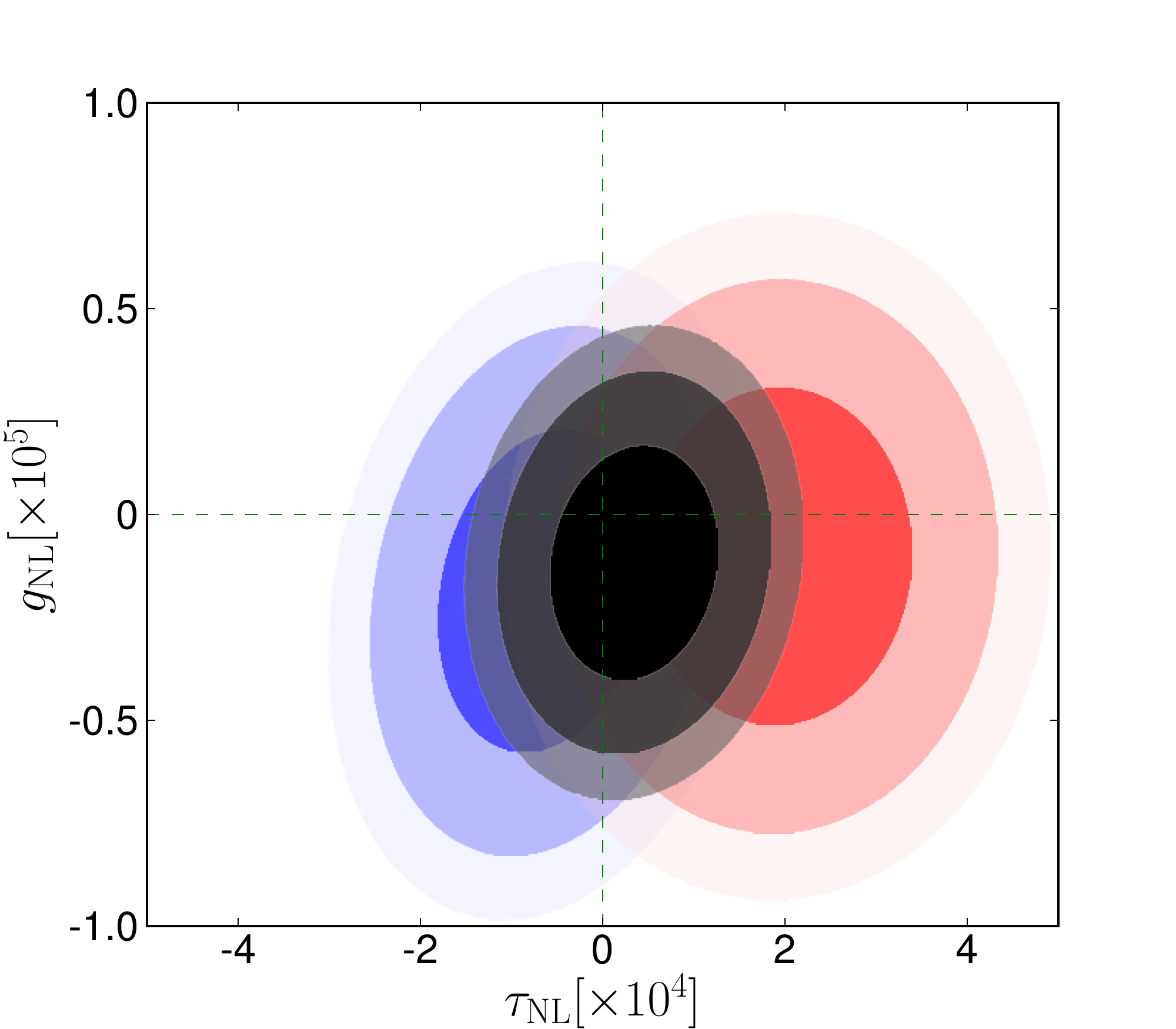}}
\caption{The 68\%, 95\%  and 99\% confidence levels for different combinations are indicated by the transparency of the contours. The frequency combinations $143\times 143$ GHz, $143\times 217$ GHz and $143\times 143 + 143\times 217$ GHz are shown in blue, red and black colors.}
\label{2D}
\end{figure}

We calculate both trispectra $K_L^{(2,2)}$ and $K_L^{(3,1)}$ from Gaussian simulations and data for Planck. The Gaussian term in the trispectra $K_L^{\rm{Gaussian}}$ is averaged from 100 Planck simulations for frequency combinations $143\times 143$ GHz, $143\times 217$ GHz and $217\times 217$ GHz, and is removed from the raw signal, which is defined as the combination of the connected part and the disconnected part. All the trispectra are shown in Fig.~\ref{RawTrispectrum}. It is seen that the disconnected components dominate the raw signal and our simulations can precisely recover these significant biases. Also, all the trispecta show consistent shapes. From 100 simulations, the full covariance matrix $\textbf{M}$ is obtained for each frequency combination and the vector $V_b=(V_b^{(2,2)},V_b^{(3,1)})$. Here $b$ is index of trispectrum band. We choose five bands for each spectrum: $L$=[2,152], [152,302], [302,452], [452,602], [602,800]. Here we use $\Delta L=150$ and $L_{\rm{cut}}=800$. We want to both avoid systematic issues with the high $L$ trispectra and get enough signal-to-noise, so we choose this conservative cut here.

We choose a binning function to maximize the sensitivity
\begin{equation}
\hat V_b=\sum_{L\in b}{w_{bL}\hat S_L}=\frac{\sum_{L\in b}{S_L\hat S_L/N^2_L}}{\sum_{L\in b}{S^2_L/N^2_L}},
\end{equation}
here $S_L=(2L+1)K_L$ is the fiducial model with $\tau_{\rm{NL}}=g_{\rm{NL}}=1$, $N_L=(2L+1)K^{\rm{Gaussian}}_L$ and $\hat S_L=(2L+1)\hat K_L$ which is the connected trispectrum from the simulation or data.

The likelihood function of the data is given  as
\begin{equation}
\chi^2(\tau_{\rm{NL}},g_{\rm{NL}})=\sum_{\nu}\sum_{bb'}(V^{(\nu)}_b-\hat V^{(\nu)}_b)M^{-1,(\nu)}_{bb'}(V^{(\nu)}_{b'}-\hat V^{(\nu)}_{b'}),
\end{equation}
where the two free parameters are $\tau_{\rm{NL}},g_{\rm{NL}}$, $b$ index of the band, and $\nu$ the index of the frequency combination.

\begin{table}
\caption{The constraints of $\tau_{\rm{NL}},g_{\rm{NL}}$ with $\Delta L=150$ and $L_{\rm{cut}}=800$ from different frequency combinations. The $68\%$ C.L. is given by $\Delta\chi^2=2.3$ except the last row.}\label{t1}
\begin{tabular}{ccc}
\hline    
$\rm{Freq. Combination}$&$\tau_{\rm{NL}}[\times 10^4]$&   $g_{\rm{NL}}[\times 10^5]$        \\        
  $143\times143$ & $-0.6\pm 1.2$ & $-1.9\pm 3.9$ \\
$143\times 217$ & $1.9\pm 1.5$ & $-1.0\pm 4.1$ \\
$143\times 143+143\times 217$ &$0.3\pm 0.9$& $-1.2\pm 2.8$ \\
$143\times 143+143\times 217$ &$0$& $-1.3\pm 1.8$ \\
  \hline  
\end{tabular}
\end{table}

\begin{table}
\caption{The constraints of $\tau_{\rm{NL}},g_{\rm{NL}}$ with different $\Delta L$ and $L_{\rm{cut}}$ for the combination $143\times 143+143\times 217$. The $68\%$ C.L. is given by $\Delta\chi^2=2.3$.}\label{t2}
\begin{tabular}{ccc}
\hline    
$143\times 143+143\times 217$&$\tau_{\rm{NL}}[\times 10^4]$&   $g_{\rm{NL}}[\times 10^5]$        \\        
$[\Delta L=150, L_{\rm{cut}}=800]$ &$0.3\pm 0.9$& $-1.2\pm 2.8$ \\
$[\Delta L=150, L_{\rm{cut}}=850]$ &$0.3\pm 0.9$& $0.3\pm 1.5$ \\
$[\Delta L=150, L_{\rm{cut}}=900]$ &$0.4\pm 0.9$& $1.7\pm 1.4$ \\
$[\Delta L=200, L_{\rm{cut}}=800]$ &$0.6\pm 0.9$& $-0.6\pm 3.0$ \\
  \hline  
\end{tabular}
\end{table}

\begin{figure}
\rotatebox{0}{\includegraphics[width=8.5cm, height=7cm]{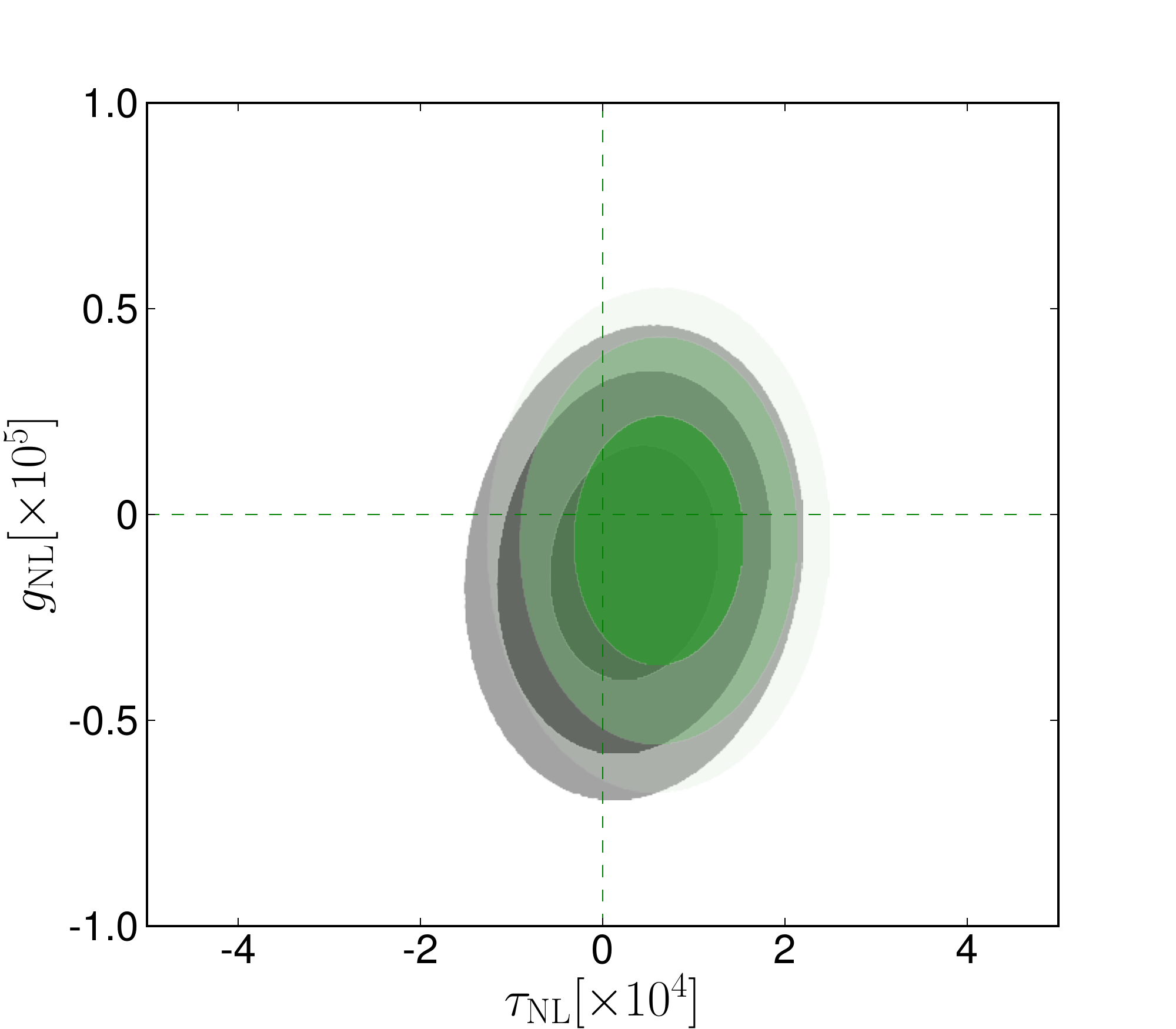}}
\rotatebox{0}{\includegraphics[width=8.5cm, height=7cm]{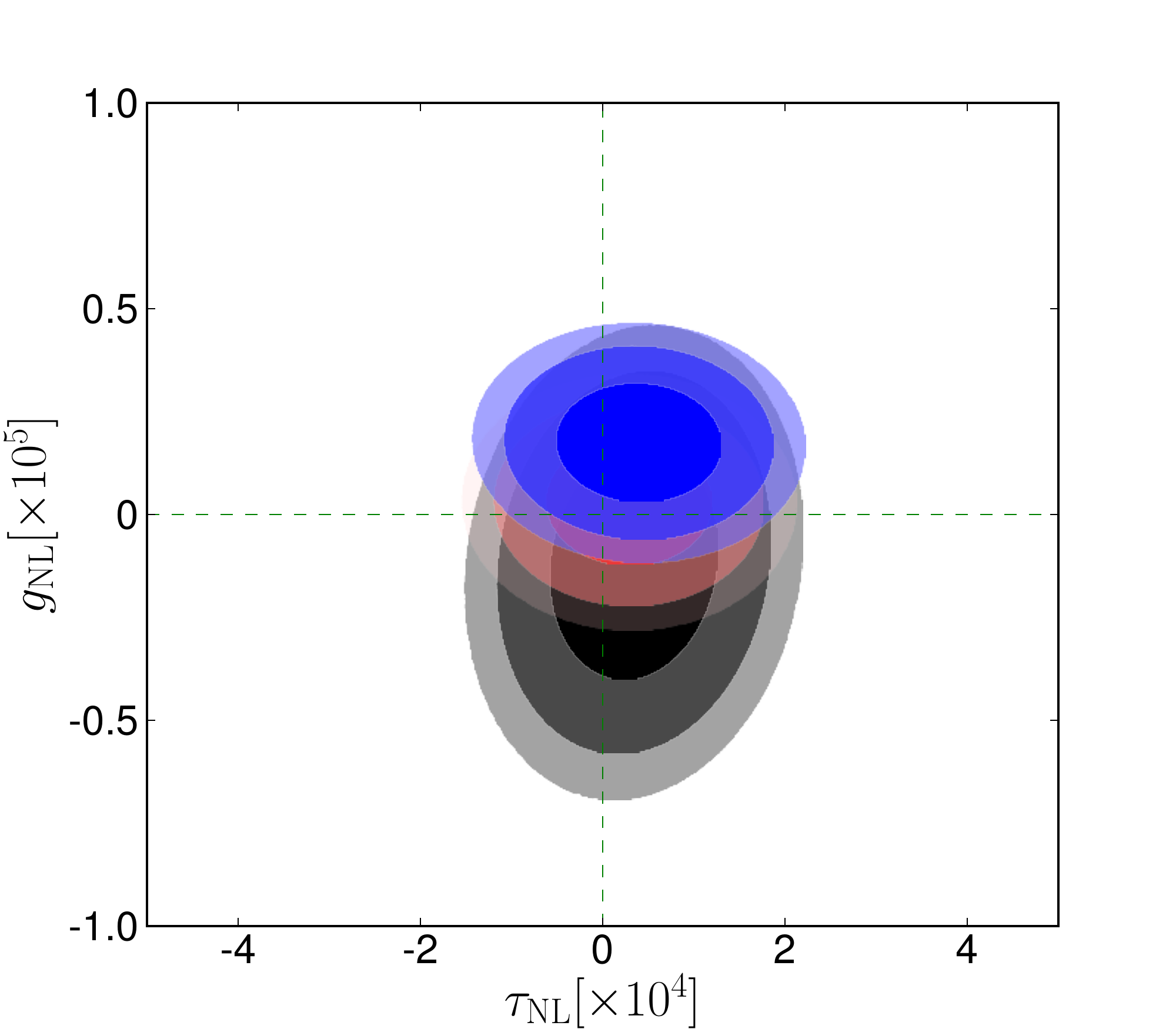}}
\caption{The 68\%, 95\%  and 99\% confidence levels for the combination $143\times 143+143\times 217$ with different bin sizes (top) and $L_{\rm{cut}}$ (bottom) 
are indicated by the transparency of the contours. In the top, for $\Delta L=150$, the contour is shown in black and green for $\Delta L=200$. For both cases, $L_{\rm{cut}}$=800.
In the bottom, $L_{\rm{cut}}=800$ is shown in black, $L_{\rm{cut}}=850$ in red, $L_{\rm{cut}}=900$ in blue. In these cases $\Delta L=150$.}
\label{2D}
\end{figure}

We draw $\mathcal{O}(10^6)$ samples for two parameters from Monte Carlo Markov chains with flat priors $-10^6\le\tau_{\rm{NL}}\le 10^6$ and $-10^7\le g_{\rm{NL}}\le 10^7$. The 217 GHz map is still significantly contaminated by CIB although we use a very conservative cut which removes $40\%$ of the sky, so we do not include $217\times 217$ GHz into our parameter estimation. The constraints for $\tau_{\rm{NL}}$ and $g_{\rm{NL}}$ are listed in Table~\ref{t1}. In the last row of Table~\ref{t1}, we show the 1-parameter constraint on $g_{\rm{NL}}$ with $\tau_{\rm{NL}}=0$. For all the combinations, we find that $\tau_{\rm{NL}}$ and $g_{\rm{NL}}$ are consistent with zero. We check the consistency between different frequency combinations in Fig.~\ref{2D}. From Fig.~\ref{2D}, it is seen that different bin sizes do not change the results. We also check the impact of effective $L$ range on the parameters. From Fig.\ref{2D}, we find that adding more $L$ range can result in a higher value of $g_{\rm{NL}}$ and the interpretation is that the high $L$ range is systematically contaminated by unresolved point sources and non-Gaussian contribution of CIB beyond the foreground mask. All the results shown in Fig.~\ref{2D} are summarized in Table~\ref{t2}.

{\it Summary:} We present the first joint constraints on $\tau_{\rm{NL}},g_{\rm{NL}}$ using Planck kurtosis power spectra that trace 
square temperature-square temperature and cubic temperature-temperature power spectra. The Gaussian biases in these statistics are
corrected for with simulations and we make use of non-Gaussian simulations to test our pipeline. We find the best joint estimate of
the two parameters to be $\tau_{\rm{NL}}=(0.3 \pm 0.9) \times 10^4$ and $g_{\rm{NL}}=(-1.2 \pm 2.8) \times 10^5$.
If $\tau_{\rm NL}=0$, $g_{\rm NL}=(-1.3 \pm 1.8) \times 10^5$.

\begin{acknowledgments}
AC and CF acknowledge support from NSF AST-1313319 and James B. Ax Family Foundation through a grant to Ax Center for Experimental Cosmology.
DR acknowledges support from the Science and Technology Facilities Council [ST/L000652/1] and from the European Research Council [ERC Grant Agreement No. 308082].
\end{acknowledgments}

\newpage

\bibliography{ng}
\end{document}